\documentstyle[bo99,epsfig]{article}

\def\ecs{\rm \,erg~ cm^{-2}\,s^{-1} }
\def\sax{{\it BeppoSAX}}

\title{Hard synchrotron BL Lacs:\\ the case of 1ES 1101-232}
\author{Anna Wolter$^1$, Gabriele Ghisellini$^1$, Gianpiero Tagliaferri$^1$, 
Fabrizio Tavecchio$^1$, Alessandro Caccianiga$^2$ }
\affil{1) Osservatorio Astronomico di Brera, Milano, Italy 2) Observat\'orio
Astron\'omico de Lisboa, Lisboa, Portugal}

\begin{document}

\maketitle

\begin{abstract}
The bright X-ray selected BL Lac object 1ES1101--232 shows a
flat X-ray spectrum, making it detectable with high statistics
over the wide BeppoSAX energy range. We have observed it in two
different epochs with BeppoSAX, and found a variation of the flux of
about 30\% that can be explained by a change in the spectral index
above the synchrotron peak. We present here the data and infer
limits on the strength of the magnetic field based on models of 
emission for High-frequency  peaked BL Lacs.

\end{abstract}

\section{Introduction}

Overall spectral energy distributions (SED) of BL Lacs 
and blazars in general show
two broad peaks: the synchrotron one at low energies and the 
inverse Compton scattering peak at high energies. The position of the 
synchrotron peak defines different classes of BL Lacs:
the HBL (High-peaked BL Lacs) and 
the LBL (Low-peaked BL Lacs). 
Ghisellini et al. (1998) and Fossati et al. (1998) propose a sequence for
blazars in which the energy of the peak is anti-correlated with the 
bolometric luminosity, 
and fainter objects, as HBL, should have a peak in the UV--X-ray band.

1ES 1101--232 (z=0.186) is an extreme case of HBL in which the
synchrotron component peaks in the X-ray band ($\sim$ 1 keV),
as shown by our previous observation (Wolter et al. 1998).
Even if not as extreme as that of the flaring states
of Mkn 501 (Pian et al. 1998) and 1ES 2344+514 (Catanese et al. 1998),
the SED of 1ES 1101--232 makes it a good candidate for TeV emission.

\section{X-ray data}

\sax\ has observed 1ES1101--232 on two occasions, on 4 Jan 97 and 19 Jun 98. 
A single power law fit with Galactic absorption at low energy is rejected
for both observations, while a broken power law yields an acceptable 
$\chi^2$. In Wolter et al. (1999) all the details of the fits 
are reported.
The broken power law model is preferred, from a statistical point of
view besides for physical 
reasons, even over a single power law with intrinsic absorption. 
The PDS observations, being so short, are not of sufficient statistical 
significance to put a real constraint on the spectrum.

The position of the break energy ($E_0$) and the slope  of the low energy
part of spectrum ($\alpha_1$) are the same in the two observations
within the errors.
On the contrary, the portion
of the 
spectrum at higher energies (i.e. above $E_0$) has changed between the 
two observations. We therefore fit the two datasets together, by using 
an appropriate model; the best fit of a broken power law model, 
in which only the high energy index $\alpha_2$ is untied between the two 
observations, is acceptable (see Table 1).

The fluxes are consistent with those obtained by the separate fits. 
Only the intensity above 2 keV changed (of $\sim 32\%$) between the two 
observations.
Even if the flux variation is small, this result might bear an impact on
spectral variability models in BL Lacs. 

\section{Spectral Energy Distribution} 

\hbox{
\vbox{\hsize=7cm
\centerline{\psfig{file=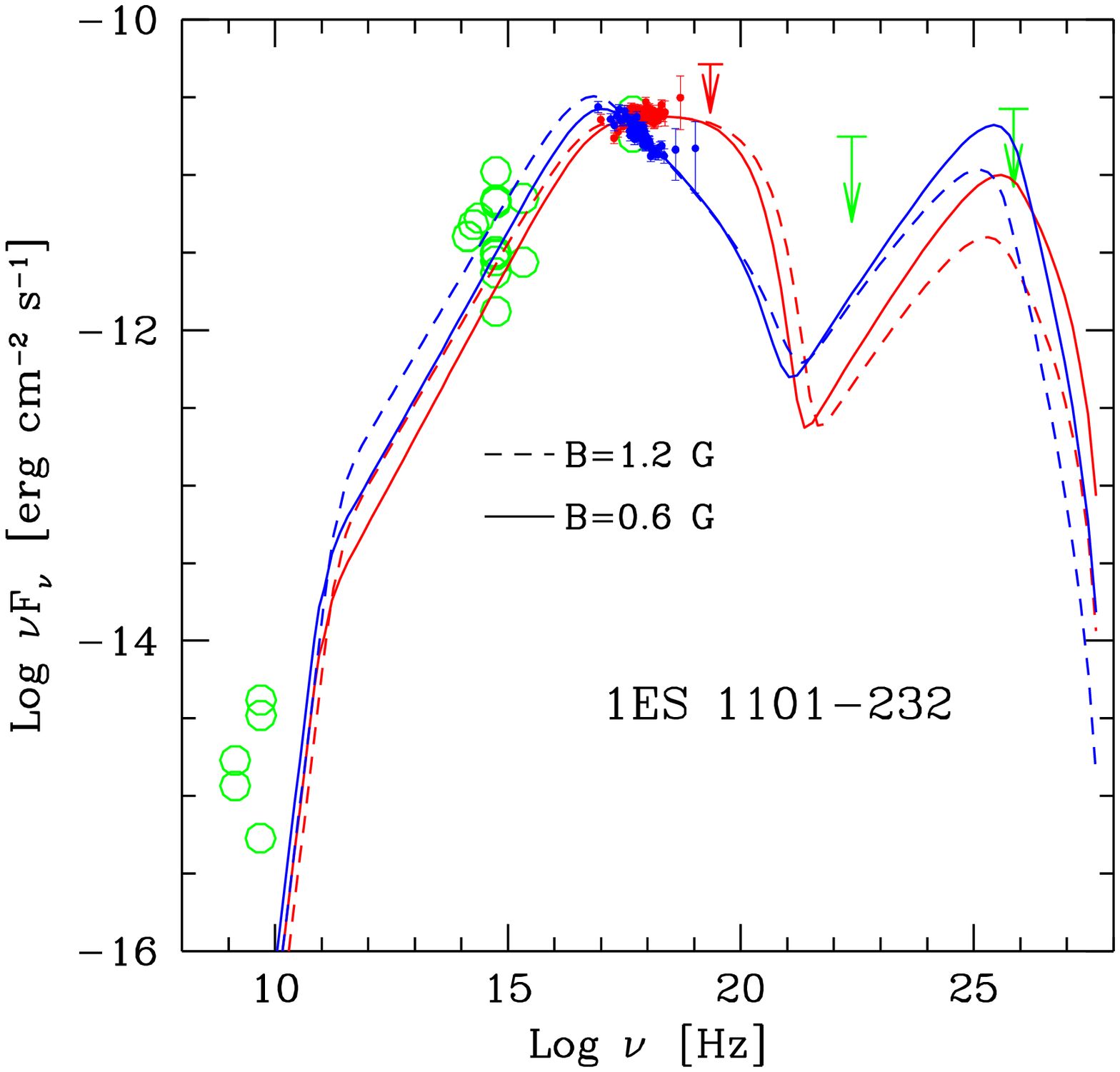,width=6.5cm} }
} \hfill
\vbox{\hsize=5cm
\small{\it
\noindent 
Figure 1: SED, points from literature and \sax\ observations. See text for 
an explanation of the model. Light gray line and dots refer to the Jan 1997
observation, while dark line and dots to the June 1998 \sax\ observation.
}
} 
} 

\vskip 0.2cm

By using the same data as reported in Wolter et al. (1998) and 
adding the second \sax\ observation we construct the SED
of Figure 1 and 2. 
Furthermore, 
1ES1101-232 has been observed on the nights of 19-27 May 1998
with the Durham University Mark 6 atmospheric $\check{\rm C}$erenkov telescope
(Chadwick et al. 1999). 
The source was not detected and an upper limit of $f_{TeV}$ ($> 300$ GeV)
= 3.7 $\times 10^{-11}$ photons cm$^{-2}$ s$^{-1}$  has been derived from the 
observation.  This value also has been plotted in Figure 1.

\begin{table*}
\caption{Broken power Law fit results for LECS+MECS data COMBINED} 
\begin{tabular}[h]{| l c r c r r |}
  \hline
Date   &  $\alpha_1$ & $\alpha_2$ & $E_0$ & F$^{\rm a}$ &$\chi^2$ (dof)  \\
       &  En. index  & En. index  & keV   &           &     \\
  \hline
Jan '97  & 0.64(0.51-0.76) & 0.97(0.93-1.03) & 1.28(1.16-1.41) &38.7 & 455.2(397)  \\
Jun '98  &   same          & 1.31(1.27-1.35) &  same           &25.5 &  \\
  \hline
\end{tabular}
\normalsize
\begin{list}{}{}
\item  Broken p.l. with $N_{\rm H}$=$N_{\rm H}^{\rm Gal}$; $\alpha_1$ and $E_0$ tied between 
the two datasets.\\
$^{\rm a}$ Unabsorbed flux [2-10 keV] in $10^{-12} \ecs$.
\end{list}
\label{fit2}
\end{table*}

We can reproduce the observed SED by using the homogeneous Synchrotron--Self 
Compton model described in detail in Ghisellini et al. (1998). 
A power-law distribution of electrons with slope $n$ and minimum Lorentz 
factor $\gamma_{min}$ is continuously injected in a
spherical region with radius $R$. The source is in relativistic motion toward
the observer and
relativistic effects are expressed by the Doppler factor $\delta $. Electrons
are free to cool and form the low energy flat spectrum with spectral index
$\alpha =0.5$.

\hbox{
\vbox{\hsize=7cm
\centerline{\psfig{file=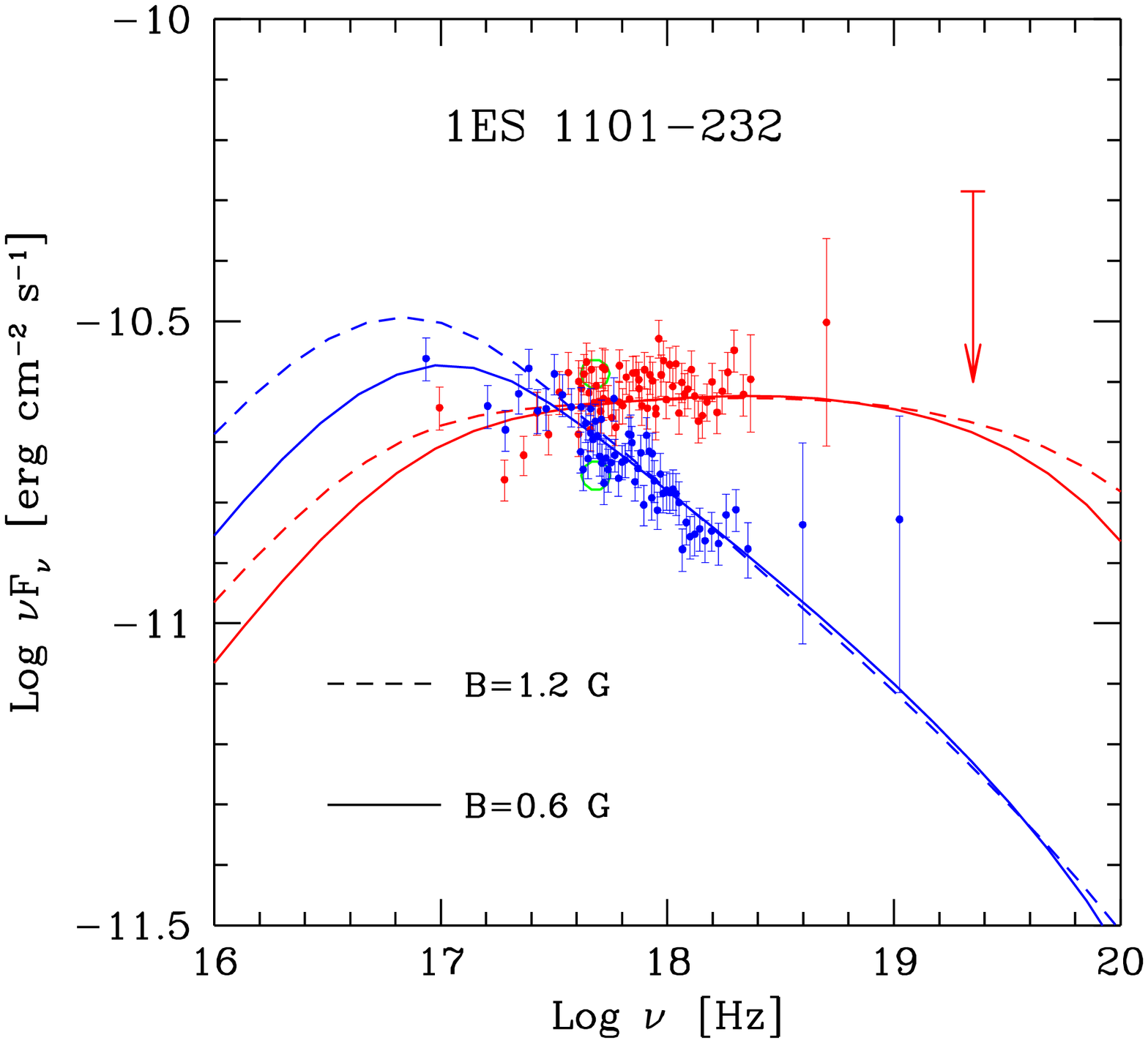,width=6.5cm} }
} \hfill
\vbox{\hsize=5cm
\small{\it
\noindent 
Figure 2: SED, enlargement of the X-ray band with the two observations and
model. Light gray line and dots refer to the Jan 1997
observation, while dark line and dots to the June 1998 \sax\ observation.
The agreement between the model and the X-ray points in the two observations
is evident.
} 
} 
} 

\smallskip
The model over-imposed on the SED is derived assuming a radius of 
$R=1\times10^{16}$ cm, $\delta$=15, 
$L_{\rm inj}^{\rm intr.}=9.3 \times 10^{41}$ erg/s; $\gamma_{max}=4 \times 10^6$,
with no external photons.
The slope of the injected electrons is $s$=2.7 (1998) or
$s$=1.95 (1997).
$B$ = 0.6 Gauss 
(and  $\gamma_{min}^{inj} = 5. \times 10^4$) for the continuous line; 
$B$ = 1.2 Gauss (and $\gamma_{min}^{inj} = 3. \times 10^4$)
for the dashed line.

\section{Magnetic field}

A small change in the magnetic field, while still consistent 
with the X-ray (\sax) observations (see Figure 1), 
produces a very different TeV emission. The TeV band data can therefore put
stringent constraints on the magnetic field.

The TeV upper limit indicates that the Compton peak cannot be higher than 
the synchrotron peak (${L_C / L_S} \leq 1$); using the analytical relations 
discussed in Tavecchio et al. (1998) we can calculate the minimum $B$ allowed 
by the observed TeV upper limit for different values of $\nu _c$ and $\delta $.
The values that produce a SED in agreement with both the X-ray spectra and the 
TeV upper limit are very
similar to those found for Mkn 501 (e.g. $\delta=15$ and $B$=0.2 G; Kataoka
et al. 1999) implying that the physical conditions of the two sources are also
quite similar.

\section{Conclusions}

The X-ray spectrum of 1ES 1101-232 is fitted by a broken 
power law (a single or an absorbed power law are not statistically acceptable)
with a break at 1.3 - 1.9 keV. 
From the first to the second observation, the spectrum varied at high
energies, becoming softer (steeper). The flux has therefore decreased
by about 32\%, in the 2-10 keV band.

The TeV observation has not yielded a detection. However, 
since the TeV
emission is largely sensitive to parameters like the magnetic field that
produces the Synchrotron emission, interesting limits can be put on this
quantity. Of course, more sensitive TeV instruments will 
produce more stringent constraints on the higher energy part of the spectrum
and therefore on the emission mechanisms. 

Multifrequency, simultaneous observations (e.g. optical, X-ray, TeV) 
will thus allow us to explain
the variability of the sources, both from the energetic and the spectral
distribution point of view. 

\begin{acknowledgements}
This work has received partial financial support from the Italian 
Space Agency and from the European Commission, TMR Programme, Research
Network Contract ERBFMRXCT96-0034 ``CERES'' 
\end{acknowledgements}

\end{document}